\documentclass[12pt]{article}
\usepackage{graphicx}

\begin{document}

\begin{center}
Basic notions of dense matter physics:applications to astronomy
\end{center}
\begin{center}
V. \v Celebonovi\'c
\end{center}
\begin{center}
Institute of Physics,Pregrevica 118,11080 Zemun-Beograd, Serbia and Montenegro
\end{center}
\begin{center}
vladan@phy.bg.ac.yu
\end{center}


\begin{abstract}
The aim of this paper is to present basic notions of dense matter physics and some of its applications to geophysics and astronomy.Topics covered in the paper include:basic observational data,fun-
damental ideas of static high pressure experiments, notions of theoretical dense matter physics, and finally some details about theoretical work on dense matter physics and its astronomical applications in Serbia.
\end{abstract}

\begin{center}
Introduction
\end{center}

Astronomy is a science full of contrasts. Nearly all physical quantities measurable in astronomy reach values which are hardly immaginable on Earth. Two interesting extremes which occur in space science concern the mass densities of the objects studied: they span about {\bf 30} orders of magnitude. At the lower limit, astronomers encounter various kinds of nebulae, the interstellar and intergalactic medium, where the density can be as low as $10^{-15}$ kg $m^{-3}$. Near the upper limit are various kinds of objects in which matter is subdued to high values of pressure (and temperature) and where the density can go upwards to values of the order of $10^{15}$ kg $m^{-3}$.

The aim of this lecture is to review the basic notions of dense matter physics and its applications in astronomy. The field is huge,and the present lecture is far from being a complete review of the field. The topics which are covered are listed in the abstract; they were chosen so as to reflect the author's research experience but combined in a way to make this lecture an introduction to the basic notions and some recent research results.
\newpage

\begin{center}
Basic data
\end{center}

Celestial objects in general, and those in the planetary system in particular, are observable since prehistoric times. The nature of space science has dramatically changed since the invention of the telescope, as attempts to measure and draw conclusions steadily gained over pure contemplation of phenomena. 

The telescope, in combination with the laws of celestial mechanics, gave the possibility of determining masses and radii of various objects. Using these data,it became possible to calculate the mean densities of objects observed.

Starting from the mass $M$ and radius $R$ of an object, and assuming spherical shape, it is clear that the volume is $V=(4/3)\pi r^{3}$, while its density is $\rho=M/V$ . This kind of calculation gave the first possibility of gaining some knowledge on the chemical composition of other planets. Densities of various planets, derived from the observed data, could be associated with various materials known on Earth. Modern values of planetary densities, taken from the National Solar System Data Center  at http://nssdc.gsfc.nasa.gov/planetary are presented graphically in Fig.1. 
Planetary densities are mutually widely different: from those which are rarer than water to those which are 5-6 times denser. This is a consequence of a set of processes in which the planetary system was formed about 4.6 billion years ago. 
\begin{figure} 
\includegraphics[width=10cm,height=7cm]{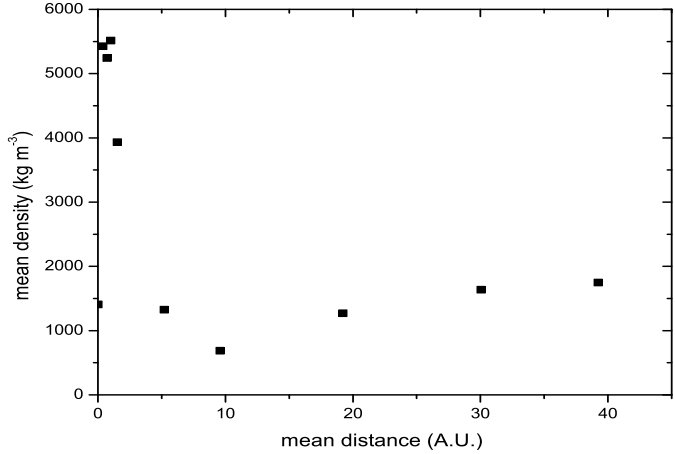}
\caption{Densities in the planetary system}
\end{figure}
It is usually taken that the present distribution of planetary densities is a consequence of the spatial distribution of temperature and chemical composition of the material in the protoplanetary disk. For an example of recent work on planetary formation see \cite{PATE:06} and  references given there. 

Real experimental work in planetary science became possible in the second half of the $XIX$ century,with the developement of seismology. The physical idea on which seismology is based is extremely simple. A point in the interior of the Earth in which an earthquake occurs becomes a source of waves which propagate along the surface and through the interior of the planet. Measurements of the period,amplitude and damping rate of the oscillations give information about the earthquake which is the source of the elastic waves, but also on the physical conditions along their path.

Figure 2 is a historic example of a seismogram \cite{WALD:93}. It shows the NS seismic waves recorded in G\"ottingen (Germany) during the earthquake in San Francisco on April 18,1906. The fact that the quake was recorded by the relatively crude equipement of the time shows the sensitivity of the method, but also the power released in this event. The symbols P and S denote the compressional and shear waves respectively,while PP and SS are waves which bounced 1 or 2 times from the surface.

\begin{figure} 
\includegraphics[width=12cm,height=8cm]{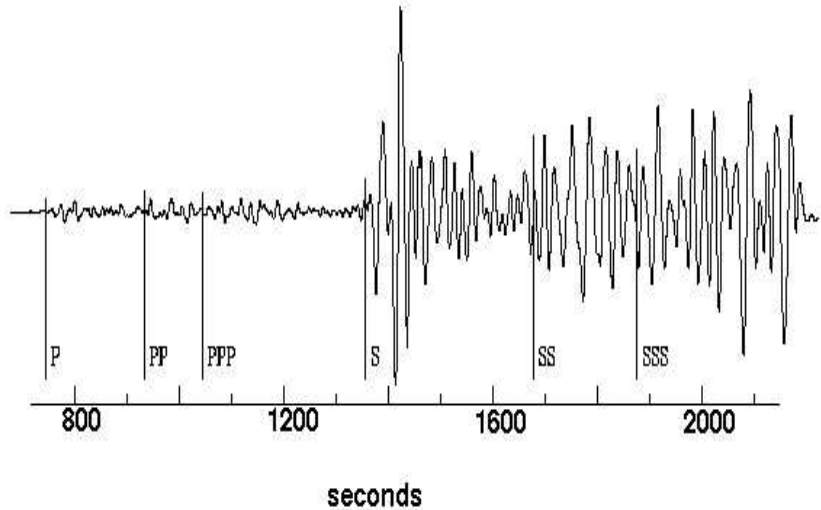}
\caption{NS waves in the San Francisco earthquake of 1906.}
\end{figure} 

The main data which can be deduced from the analysis of seismograms are the values of the speed of propagation of different waves.  Analyzing the wave profiles gives the possibility of determining the position of various discontinuities deep in the core \cite{COST:05}. The measured values of the speeds of propagation open the possibility of determining  the compressibility of the mixture of materials through which the wave propagates. It is known that in a liquid the speeed of a compressional wave is related to the adiabatic compressibility $K_{S} $ in a simple way:
\begin{equation}
	v_{P}^{2}= \frac{K_{S}}{\rho}
\end{equation}
where
\begin{equation}
	K_{S}=\frac{1}{\rho} \frac{\partial \rho}{\partial P}
\end{equation}
The logic behind the applicability of these two expressions in planetology is simple: in the case of the Earth,it is usually assumed that its outer core is  liquid. Accordingly,by measuring the speed of seismic waves resulting from an earthquake,eqs. (1) and (2) give the possibility of drawing some conclusions concerning the equation of state of the material. These expessions are based on a simplified model; note that real work in seismology is based on more complicated equations,but the reasoning behind remains the same. 
\begin{center}
Laboratory experiments: a few notes
\end{center}
  
Historians of science have discovered that a certain wealthy English gentleman named Mr.Canton sometime in the $XVIII$ century made a first attempt to perform an experiment under high static pressure. He compressed water at room temperature to a pressure of the order of 0.1 GPa. To his astonishement water was transformed into ice \cite{BLOCK:80}. We know today that the phase diagram of ice under high pressure has at least 12 phases \cite{HOLZ:99}.  

Systematic experimental work on materials under high pressure has been initiated towards the end of the $XIX$ century by P.W.Bridgman at Harvard University \cite{BRIDG:64}. For his life-long work Bridgman was awarded the Nobel prize in physics in 1946. In his experiments, Bridgman used presses which could contain large samples and had small $P-T$ gradients,which was very useful for the measurement process. At the same time, the accessible region of the $P-T$ plane was extremely small, and the presses were expensive to build and complicated to operate.

A big breakthrough occured in the middle of the last century with the invention of the diamond anvil cell (DAC). Details about the construction of this instrument,the choice of diamonds and some of the experimental methods  are avaliable in the literature such as \cite{CHE:04}. An interesting  step forward  has occured a short time ago: the so called "ruby scale" which is used in nearly all measurements in DACs has been corrected, and the validity of this modification tested to $P\leq100$ GPa,$T\leq850$ K \cite{GONCH:05}. 

\begin{center}
Examples of experimental studies
\end{center}
Drawing conclusions about the behaviour under high pressure of astronomically interesting materials is comlicated by the experimental methodology and the complexity of phenomena occuring with the increase of the density. An excellent example,spanning more than $70$ years of research and still unsolved,is offered by the behaviour of hydrogen under high pressure. 

The problem of dense hydrogen can be resumed in the following way. It has been predicted numerous times in the last 70 years that hydrogen should become metallic at a pressure of the order of $250-300$ GPa. These predictions were made by different authors and theoretical methods. As they all were giving the same order of magnitude,it was taken almost as an experimental result. However,when such values of pressure became accessible,and experiments performed \cite{NARA:98},hydrogen {\it did not} become a metal for $P\leq342$ GPa. More recent experiments in a DAC have shown that molecular hydrogen continues to exist up to $P=316$ GPa and that it becomes black at $P=320$GPa \cite{LOUB:02}. It was also experimentally obtained that the $semiconductor\rightarrow metal$ transition in fluid hydrogen occurs at $P=140$ GPa; $T=3000$ K \cite{NELL:98}. Finally,note that recent theoretical work indicates the possibility that hydrogen turns into a metal only at $P\geq400$GPa \cite{BABA:05}. 

In the field of shock compression experiments,the situation also seems to be stationary. At the time of this writing (October ,2005) there are no new results from the group in the Lawrence Livermore National Laboratory (LLNL) performing laser shock experiments. Some experimental work is going on in Rochester,but the next opportunity to obtain " big" results will have to wait for the completion of the National Ignition Facility (NIF). This facility is being built in LLNL,and it is expected to be completed in ~2008.In the meantime " flying plate"  experiments are being performed in Sandia Labs. (USA) and in Arzamas (Russia). They have reached 1.4 MBar,but their results for the equation of state (EOS) of hydrogen do not agree with previous shock compression results from Livermore \cite{DW:05}. Space limitations do not allow going into further details, but the point is that although hydrogen is the simplest chemical element,the most abundant in the Universe,its high pressure behaviour is extremely hard to predict.      

Another closely related material is water.Its importance for the existence of life on Earth is clear. In planetology,its importance stems from the fact that water ice (or ices containing water and different other materials) exist on some planets and their satellites and as such contributes to their albedos. This implies that knowledge of the phase diagram of ice(s) is necessary for the correct interpretation of the observed values of albedos of various planetary system objects. The "complexity"  arises from the fact that the phase diagram of ice has {\bf at least} 11 parts. The physics of ice is of considerable importance in studies of interstellar grains.See,for example,\cite{VBR:05} and references given there).  

Studies of materials under high pressure are important for geophysics. Just as an example,note a recent experiment in which the ellasticity of hcp iron has been studied in a DAC at room temperature for $P=112$ GPa. The anisotropy of the velocity of sound has been measured,and the result is in close correspondence with the anisotropy of the Earth's inner core \cite{ANTO:04}. 

\begin{center}
A note on theoretical work
\end{center}

Technological developement has made possible the enlargement of the experimentally accessible part of the $P-T$ plane. However,an incomparably greater region of the same plane can be studied theoretically. Any specimen of a real material represents a typical example of a many body problem.In rigorous terminology of statistical physics,the {\it many body} problem is defined as the eigenproblem of the following Hamiltonian
\begin{equation}
H=\sum_{i=1}^{N}(-\frac{\hbar^{2}}{2m})\nabla_{i}^{2}+\sum_{i=1}^{N}V(\vec{x}_{i})+\sum_{i,j=1}^{N}v(\vec{x}_{i}-\vec{x}_{j})
\end{equation}

In this equation {\it N} denotes the number of particles in the system,{\it m} is their mass and all the other symbols have their usual meanings. In all real examples of many-body systems,the sums in this expression are {\it impossible} to calculate exactly,because the number of particles {\it N} is of the order of $10^{23}$. The art of theoretical dense matter physics in fact consists in finding methods of calculation of sums in eq.(3),because this ultimately leads to the possibility of determination of the thermodynamical potentials (and phase transition points) of a system. Attempting to review the whole field would be out of place,so it the following we shall concetrate only on a contribution to dense matter physics by researchers in this country.

\begin{center}
The contribution from this country
\end{center}

At the beginning of the sixties,Pavle Savi\'c and Radivoje Ka\v sanin have started developing a semiclassical theory of dense matter.Their theory is called "semiclassical"  because it uses some notions of classical in combination with some ideas of atomic physics. The starting point of their research was trying to find a simple relationship between the solar and planetary mean densities,starting form the observed data known at the time. It turned out that such a relationship exists and has the following form:
\begin{equation}
\rho=\rho_{0} 2^{\varphi}
\end{equation}
where $\rho_{0}$ is the mean solar density and $\varphi$ is an integer. Choosing $\varphi\in[-2,2]$ it becomes possible to reproduce the observed planetary densities. Although this fit "works" Savi\'c and Ka\v sanin did not propose a physical mechanism which could be at its origin. Using this result and a known fact from both geophysics and laboratory high pressure experiments that at some values of pressure abrupt changes of the density of materials occur,they qualitatively concluded that the electronic structure of materials changes under high pressure \cite{SAKA:62}. 

For the sake of historical completeness,it should be noted that they were not the first to propose such an idea;the first to advance the idea of "atomic destruction" under high external pressure was P.W.Bridgman back in $1927$ \cite{ASHC:04}. A couple of years after,Fermi played with the same idea,while exploring various possibilities offered by the (then new) Schr\"odinger equation.  

Starting from these ideas,Savi\'c and Ka\v sanin proposed a set of 6 experimentally founded postulates which govern the behaviour of materials under high pressure. Developing these postulates,they have set up a calculational scheme,which gives the possibility of theoretical studies of dense matter physics,both in laboratory and astronomical applications. Details are given in (for example) \cite{CHE:00}.

Input data needed for the modelling of the internal structure of a planet,

satellite or asteroid within this theory are only the mass and the radius of the object. Starting from this couple of values,it becomes possible to calulate the values of the following parameters: 
\begin{itemize}
	\item the number of layers in the interior and their thickness;
	
	\item the distribution of $P$, $\rho$, $T$ within the layers;
	
	\item the mean molecular mass of the chemical mixture that the object is made of;
	
	\item the magnetic moment of the object;
	
	\item the limits of the physically allowed interval of values of the  angular frequency of rotation of the object.
\end{itemize}

All planets except Saturn and Pluto have been modelled. Models of the Moon,
Galileian satellites,the satellites of Uranus, Triton and Titan,as well as the asteroids 1 Ceres and 10 Hygiea have been calculated so far. The first celestial body to be modelled within this theory was the Earth;as an illustration,table 1 contains the main results on the interior of our planet.

\begin{table}
\begin{center}
\begin{tabular}{|c|c|c|c|c|}
\hline
depth$[km]$& 0-39 & 39-2900 & 2900-4980 & 4980-6371 \\
$\rho_{max}$[kg $m^{-3}$] & 3000 & 6000 & 12000 & 19740 \\
$P_{max}$[GPa]& 25 & 129 & 289 & 370\\
$T_{max}$[K]& 1300 & 2700 & 4100 & 7000\\
\hline
\end{tabular}
\caption{The interior of the Earth}
\end{center}
\end{table}

The algorithm used in the theory proposed by Savi\'c and Ka\v sanin is extremely simple.In spite of that,results of its applications to planetological problems are in good agreement with more complex models.A model of the crust,called $CRUST 5.1$, used by the U.S.Geological Survey, is acessible at the following internet address http://quake.wr.usgs.gov/research/structure/
CrustalStructure/index.html. In most parts of the Earth,values of the thickness of the crust are in good agreement with those given by the model of Savi\'c and Ka\v sanin. Disagreement is obvious near the poles,but finding its explanation is at present an open question. 

For all the planetary objects modelled within this theory,the value of the mean atomic mass $A$ of the material they are made of was also calculated.These values,collected from various papers,are presented in table 2. 

\begin{table}
\begin{center}
\begin{tabular}{|c|c|c|c|}
\hline
object & mean atomic mass $A$ & satellite& mean atomic mass $A$\\
\hline
Sun &1.4& Moon&71 \\
Mercury &113&J1&70\\
Venus &28.12&J2&71\\
Earth &26.56&J3&18\\
Mars&69&J4&19\\
1 Ceres&96&U1&38\\
Jupiter &1.55&U2&43\\
Saturn&/&U3&44\\
Uranus&6.5&U4&32\\
Neptune&7.26&U5&32\\
Pluto&/&Triton&67\\
\hline
\end{tabular}
\caption{The composition of the planetary system}
\end{center}
\end{table}
Interesting conclusions can be drawn from table 2. Triton and Neptune are not made up of the same material, which agrees with known results of celestial mechanics that Triton is a captured body. Differences of $A$ are also visible for the Earth-Moon system,but their explanation is more complex. Modern work on the origin of the Moon indicates that it is a result of the colision of a body of the size of Mars with the Earth, which happened about 4.5 billion of years ago.The necessary condition for the formation of the Moon is that the impact produces solid and liquid debris. Debris from this impact formed a disc around the Earth that coalesced to form the Moon. Recent numerical simulations \cite{WA:06} show that the colision was slow,with the impactor moving with a speed less than about $15$ $km s^{-1}$. According to these results,the deep interior of the Moon  contains "the remnants" of the impactor,while the surface layers are in fact debris from the impact. Judging by their values of $A$,Mars and Triton appear to be chemically similar. Using only this similarity, one could immagine that Triton formed near the present orbit of Mars,was somehow ejected from this region and captured by Neptune. This idea demands "fine tuning" of the relative velocities of all the objects in the problem,which is difficult to achieve. The relative differences of the average atomic masses are visibly large,and they are certainly mostly caused by real physical differences between various objects but to some extent also by the extreme simplicity of the theory.

A few words are in order on laboratory applications of this theory. A detailed algorithm has been developed for the calculation of the pressure at which a first order phase transition can be expected in a material under high pressure. It has been applied to about 20 materials,chosen at random,for which reliable experimental data could easily be found in the literature \cite{CHE:92}. It was shown that discrepancies with experimental results exist,but that they vary between practically zero and $~30$ percent,and that they can be explained by differences in the structure of the materials studied and the nature of the inter and intramolecular potentials in them. These results are interesting on their own,and they also lend weight to all astronomical applications of this theory. 

\begin{center}
Conclusions
\end{center}

Dense matter physics is a fascinating field of pure physics with interesting applications in space science.It is practiced in this country for several decades,and the results are very encouraging. The aim of this contribution was to review the basic notions and some results of dense matter physics in general,and obtained in this country in particular.    

\begin{center} 
Acknowledgement
\end{center}
 
I am grateful to Hugh Dewitt from the Lawrence Livermore National Laboratory for helpful information on experiments on hydrogen.


\begin{thebibliography}{99}














\bibitem{PATE:06} Papaloizou,J.C.B.and Terquem,C. {\it Rep.Progr.Phys.},{\bf 69},119 (2006).

\bibitem{WALD:93} Wald,D.J.,Kanamori,H.and Helmberger,D.V.{\it Bull.Seism.Soc.Am.},

{\bf 83},981 (1993).

\bibitem{COST:05} Cormier,V.F.and Stroujkova,A.:{\it Earth and Planet.Sci.Lett.},{\bf 236},96 (2005).

\bibitem{BLOCK:80} Block,S.,Piermarini,G.and Munro,R.G.: {\it La Recherche} ,{\bf 11},806 (1980). 

\bibitem{HOLZ:99} Holzapfel,W.B.: {\it Physica} {\bf B265},113 (1999). 

\bibitem{BRIDG:64} Bridgman,P.W.: Collected Experimental Papers ,{\bf 1-8}, Harvard Univ.Press,Cambridge,Mass (1964). 

\bibitem{CHE:04} \v Celebonovi\'c,V.: in {\it Equation-of-state and Phase-Transition issues in Models of Ordinary Astrophysical Matter},eds.V.\v Celebonovi\'c,W.Dappen and D.O.Gough,AIP Conf.Proc.Series,{\bf 731},280 (2004).

\bibitem{GONCH:05}Goncharov, A. F.,Zaug, J. M.,Crowhurst, J.C. Gregoryanz, E.: {\it J.Appl.Phys.},{\bf 97},npag.(2005).

\bibitem{NARA:98} Narayana,C.,Luo,H.,Orloff,J.Ruoff,A.L.: {\it Nature},{\bf 393},46 (1998). 

\bibitem{LOUB:02}Loubeyre,P.,Occelli,F. LeToullec,R.: {\it Nature} {\bf 416},613 2002.

\bibitem{NELL:98} Nellis,W.J.,Louis,A.A. Ashcroft,N.W.: {\it Phil.Trans.R.Soc.},{\bf A 356},119 (1998).

\bibitem{BABA:05} Babaev,E.,Sudbo,A. and Ashcroft,N.W.: {\it Phys.Rev.Lett.},{\bf 95},105301 2005.

\bibitem{DW:05}Dewitt,H.: private communication dated October 25,2005.

\bibitem{VBR:05} van Broekhuizen,F.A.: Doctoral Thesis,Leiden University (2005).

\bibitem{ANTO:04} Antonageli,D.,Occelli,F.,Roquardt et.al.:{\it Earth and Planetary Sci.Lett.},{\bf 225},243 (2004). 

\bibitem{SAKA:62}Savi\'c,P., Ka\v sanin,R.:{\it The behaviour of materials under high pressure},{\bf 1-4},Ed.by SASA:Belgrade (1962/65).

\bibitem{ASHC:04}Ashcroft,N.W.: {\it J.Phys.:Condens.Matt.},{\bf 16},S945 (2004).

\bibitem{CHE:00}\v Celebonovi\'c,V.:  {\it Publ.Obs.Astron.Belgrade} ,{\bf 67},19 (2000).



\bibitem{WA:06}Wada,K.,Kokubo,E.and Makino,J.: {\it Astrophys.J.},{\bf 638} ,in press (2006). 

\bibitem{CHE:92} \v Celebonovi\'c,V.: {\it Earth,Moon and Planets},{\bf 58},203 (1992).






\end{thebibliography}
\end{document}